\documentclass[aps,prb,twocolumn,showpacs]{revtex4-1}
\usepackage{amsmath}
\usepackage{amssymb}
\usepackage{graphicx}
\usepackage{times}
\usepackage{color}
\usepackage{subfigure}
\usepackage{setspace}
\usepackage{bm}

\makeatletter

\begin{document}

\newcommand {\beq} {\begin{equation}}
\newcommand {\eeq} {\end{equation}}
\newcommand {\bqa} {\begin{eqnarray}}
\newcommand {\eqa} {\end{eqnarray}}
\newcommand {\ba} {\ensuremath{b^\dagger}}
\newcommand {\Ma} {\ensuremath{M^\dagger}}
\newcommand {\psia} {\ensuremath{\psi^\dagger}}
\newcommand {\psita} {\ensuremath{\tilde{\psi}^\dagger}}
\newcommand {\lp} {\ensuremath{{\lambda '}}}
\newcommand {\can} {\ensuremath{{\cal S}}}
\newcommand {\Q} {\ensuremath{{\bf Q}}}
\newcommand {\kk} {\ensuremath{{\bf k}}}
\newcommand {\qq} {\ensuremath{{\bf q}}}
\newcommand {\kp} {\ensuremath{{\bf k'}}}
\newcommand {\rr} {\ensuremath{{\bf r}}}
\newcommand {\rp} {\ensuremath{{\bf r'}}}
\newcommand {\ep} {\ensuremath{\epsilon}}
\newcommand {\nbr} {\ensuremath{\langle ij \rangle}}
\newcommand {\no} {\nonumber}
\newcommand {\up} {\ensuremath{\uparrow}}
\newcommand {\dn} {\ensuremath{\downarrow}}
\newcommand {\rcol} {\textcolor{red}}
\newcommand {\bcol} {\textcolor{blue}}

\begin{abstract}
Braiding of non-Abelian Majorana anyons is a first step towards using them in quantum computing. We propose a  protocol for braiding Majorana zero modes formed at the edges of nanowires with strong spin orbit coupling and proximity induced superconductivity. Our protocol uses high frequency virtual tunneling between the ends of the nanowires in a tri-junction, which leads to an effective low frequency coarse grained dynamics for the system, to perform the braid. The braiding operation is immune to amplitude noise in the drives, and depends only on relative phase between the drives, which can be controlled by usual phase locking techniques. We also show how a phase gate, which is necessary for universal quantum computation, can be implemented with our protocol.
\end{abstract}
\title{Braids and phase gates through high-frequency virtual tunneling of Majorana Zero Modes}
\author{Pranay Gorantla and Rajdeep Sensarma}
\affiliation{Department of Theoretical Physics, Tata Institute of Fundamental
Research, Mumbai 400005, India.}

\pacs{}
\date{\today}

\maketitle

\section{Introduction}

Gapped two dimensional quantum many body systems can sustain
localized excitations (anyons) which have non-Abelian mutual
statistics~\cite{Leinaas1977,Wikczek1982,Nayak2008}. Moving a non-Abelian anyon
around another adiabatically does not result in the multiplication of the wavefunction by $\pm 1$
(bosons/fermions), or by a phase (Abelian anyons~\cite{Halperin1984,Wilczek1984}); 
rather it generates a non-trivial unitary rotation in a degenerate subspace of the
system. This process, called braiding of anyons~\cite{Ivanov2001}, is immune to local
perturbations. Braiding of anyons is a
key step towards achieving fault tolerant quantum gates, which are the
building blocks of an architecture for fault tolerant quantum 
computing~\cite{Shor1996,Preskill1998,Kitaev2003,Nayak2008}.  

Non-Abelian anyonic excitations arise either in cores of topological defects 
in ordered states~\cite{ReadGreen2000}, or as localized excitations in
strongly interacting systems~\cite{MooreRead1991}. A
promising candidate for experimental realization and manipulation of
anyons are the localized Majorana zero modes (MZMs) in
semiconductor-superconductor heterostructures with strong spin orbit coupling
~\cite{Oreg2010,SauLutchyn2010,Multiple2010,Alicea2010,Stanescu2011,Sarma2015}. 
While there are several proposals
~\cite{Oreg2010,SauLutchyn2010,Multiple2010,Alicea2010} to realize these
excitations, in recent years, MZMs have been experimentally
realized at the end points of semiconductor nanowires
with strong spin orbit coupling and proximity induced
superconductivity~\cite{MarcusSauLutchyn2016,Suominen2017,Kouwenhoven2012,Das2012,Deng2012,Churchill2013,Finck2013,Zhang2016,Marcus2016,Chen2017,Zhang2017}.


Several proposals exist in the literature for braiding of MZMs
~\cite{HellFlensbergLeijnse2017,Alicea2011coldatom,Kraus2013,AkhmerovBeenakker2012,Alicea2011,SauTewari2011,Karzig2015,Jelena2016,Multiple2017}. 
Most proposals require a tri-junction of three Majorana nanowires, which represent the
simplest 2D network of Kitaev chains~\cite{Alicea2011}.
The tri-junction can host unpaired MZMs at four
possible positions, three at the ends of the wires far from each
other, and one at the junction. In the initial and final state, two of
the wires are in the topological phase, with the MZMs at their
endpoints forming the qubit. The braiding sequence involves driving
different wires in and out of topological phases to
move the MZMs around each other~\cite{Alicea2011}. A pair of 
Majorana fermions can be linearly combined to form a complex fermion, 
and braiding corresponds to a unitary rotation by $\pi/2$ in the two dimensional 
degenerate subspace of this fermion. However, braiding alone is 
not enough to achieve universal quantum computation, and needs 
to be supplemented by a \emph{phase gate}, which corresponds to a $\pi/4$ 
rotation in the degenerate subspace mentioned above. While this operation 
is not topologically protected, it is important to obtain protocols for 
the phase gate which are immune to different kinds of noises inherently 
present in a realistic system.

In this paper, we propose a  protocol for braiding MZMs
using a sequence of high frequency tunneling drives between
the end-sites of the Majorana nanowires, together with three
ancillary Majorana operators which are not part of the qubit. The frequencies are
detuned from the pairing gap, leading to an effective slow non-dissipative dynamics in
the system, which can be treated within a high frequency
expansion. This effective dynamics generates the braiding of the
MZMs. A key feature of this protocol is that braiding is completely immune
to amplitude noise in the drives. The protocol depends only
on the phase difference between the drives, which allows the protocol
to be operated at low phase noise by locking the phase of the
drives. We extend our protocol to obtain phase gates which are immune 
to amplitude and phase noises in the drives. 

%
\begin{figure*}[t]
\centering
\includegraphics[width =0.9\textwidth]{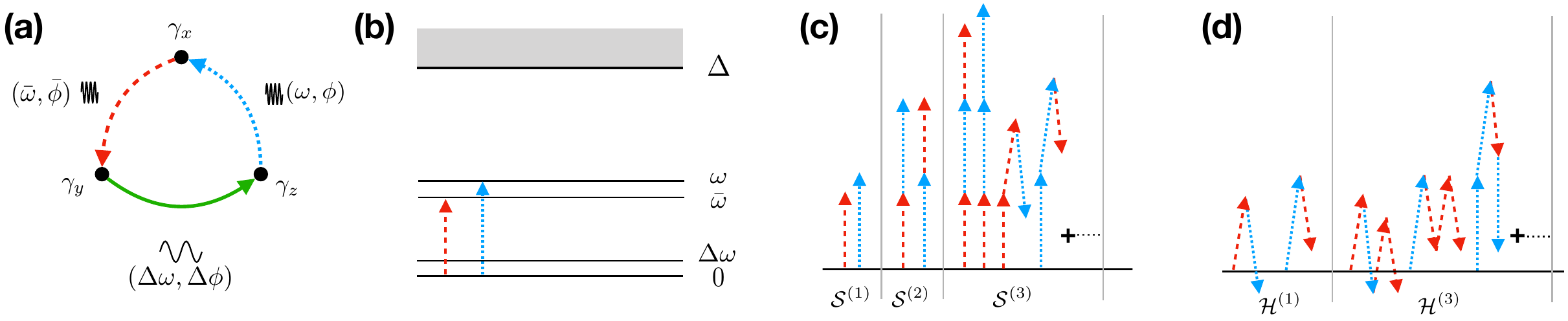}
\caption{(a) The building block of our protocol. The arrowed lines represent the couplings between Majorana operators: dashed red and dotted blue lines are the high frequency drives, and solid green line is the effective low frequency coupling. The direction of arrow represents the operator structure in the Hamiltonian, e.g., the blue dotted arrow goes from $\gamma_z$ to $\gamma_x$ and the corresponding term in the Hamiltonian is $h \sin(\omega t+\phi)\ i\gamma_z\gamma_x$. (b) Typical frequency and energy scales in the problem. $\Delta$ is the superconducting gap in the nanowire which sets the highest scale. The frequencies of dashed red and dotted blue arrows are chosen to lie deep within the gap, $\omega,\bar{\omega}\ll\Delta$, and to have a very small difference, $\Delta\omega\ll\omega,\bar{\omega}$. 
(c) Some representative terms in $\can^{(1)}$, $\can^{(2)}$ and $\can^{(3)}$ with frequencies $\sim \omega$ or higher. (d) Some representative terms in $\mathcal{H}^{(1)}$ and $\mathcal{H}^{(3)}$ with frequencies $\sim \Delta \omega$. In $\mathcal{H}$, we require even number of arrows which is possible only at odd orders.}
\label{fig:buildblock}
\end{figure*}

\section{High-frequency virtual tunneling}

We first describe a protocol to adiabatically drive the tunneling between
two Majorana operators by coupling each of them to a third 
with high frequency drives. This process is shown in
Fig~\ref{fig:buildblock}(a), where $\gamma_{x,y,z}$ are three Majorana operators 
at the end of three nanowires. We drive the tunneling 
between $\gamma_x$ and $\gamma_y$ with a
frequency $\bar{\omega}$, an amplitude $\bar{h}$, and a phase
$\bar{\phi}$, shown by the dashed red line. Simultaneously, we 
drive the tunneling between $\gamma_x$ and $\gamma_z$ with a
frequency $\omega$, an amplitude $h$, and a phase
$\phi$, shown by the dotted blue line. If the protocol is arranged
such that $|\Delta \omega| =|\omega-\bar{\omega}|, h, \bar{h} \ll
\omega, \bar{\omega}$, the separation of scales leads to an effective
description of the dynamics of the system on a scale $\sim 1/\Delta
\omega$, obtained by a high frequency expansion, which coarse grains
the dynamics on a scale $\sim 1/\omega$. We note that $\omega, \bar{\omega}$ 
must be far detuned from the superconducting gap $\Delta$, so that the system does 
not absorb and make transitions. Fig.~\ref{fig:buildblock}(b) shows a schematic of the energy 
scales involved in the process with $|\Delta\omega|\ll\omega,\bar{\omega}\ll\Delta$. For
experimentally observed $\Delta \sim 60 \text{ GHz}$
~\cite{Kouwenhoven2012}, the choice of parameters $\omega,\bar{\omega} \sim
12 \text{ GHz}$, $h,\bar{h} \sim 5 \text{ GHz}$, and $\Delta\omega \sim 1 \text{ GHz}$
is reasonable and feasible \footnote{In practice, all the drives have a finite bandwidth. For 
$\Delta\omega \sim 1 \text{ GHz}$, a bandwidth of $\sim 1 \text{ MHz}$ is easily attainable}. 
This estimate shows that our protocol is orders of magnitude faster than existing adiabatic protocols.
The key result is that, in the coarse-grained description,
$(\gamma_x,\gamma_y)$ and $(\gamma_x,\gamma_z)$ are decoupled, while
$(\gamma_y,\gamma_z)$ are coupled by an effective drive of
amplitude $\sim h^2/\omega$, modulated at $\Delta \omega$, with phase $\sim \phi-\bar{\phi}$, as
shown by the solid green line in Fig~\ref{fig:buildblock}(a).

The Hamiltonian driving $(\gamma_x,\gamma_y)$ and $(\gamma_x,\gamma_z)$ is  
\bqa
H\left(t\right)=\bar{h}\sin\left(\bar{\omega}t+\bar{\phi}\right)i\gamma_{x}\gamma_{y}+h\sin\left(\omega t+\phi\right)i\gamma_{z}\gamma_{x}.
\label{eq:majorana_drive_ham}
\eqa
In the high frequency expansion, the
inverse propagator $G^{-1}=i\partial_t- H(t)$ is transformed by a
unitary operator $e^{-i {\cal S}(t)}$ to $\mathcal{G}^{-1}=e^{i
\can(t)}G^{-1}e^{-i \can(t)}=i\partial_t -\mathcal{H}(t)$. Using
$[i\can,i\partial_t]=\dot{\can}$, we get
\beq
\mathcal{H} =\sum_{m=0}^{\infty} \frac{1}{m!}\bigg[i\can,\bigg[i\can, ...\bigg[i\can,H -\frac{\dot{\can}}{m+1}\bigg]...\bigg]\bigg].
\label{eq:heff_comm}
\eeq
Here, $\can$ and $\mathcal{H}$ have high frequency expansions, i.e., 
$\can =\sum_{n=1}^\infty \can^{(n)}(t)$ 
and $\mathcal{H} =\sum_{n=0}^\infty \mathcal{H}^{(n)}(t)$, where 
$\can^{(n)} \sim (h/\omega)^n$ and $\mathcal{H}^{(n)}(t) \sim h^{n+1}/\omega^{n}$. 
$\can^{(n+1)}$ is determined in such a way
that $\mathcal{H}^{(n)}$ has only low frequency terms [static
or ${\cal O}( \Delta\omega)$], while $\can$ contains only high
frequency terms $\sim \omega$ or higher harmonics. Fig.~\ref{fig:buildblock}(c) 
shows some representative terms for $\can^{(1)}$, $\can^{(2)}$, 
and $\can^{(3)}$ while Fig.~\ref{fig:buildblock}(d) shows similar terms for $\mathcal{H}$. 
Each term in the figures is a sequence of dashed red and dotted blue arrows 
which correspond to drive terms with frequencies $\bar{\omega}$ and $\omega$ respectively.
It is clear that an even number of arrows is required to create terms 
modulated at frequencies $\sim \Delta\omega$, and hence, $\mathcal{H}^{(n)}$ 
is non-zero only for odd $n$. At order $2p+1$ (p is non-negative integer), $\mathcal{H}^{(2p+1)}$
is a product of terms $\sim (\gamma_x\gamma_y)^k (\gamma_x\gamma_z)^{2p+2-k}\sim
\gamma_y^k\gamma_z^{2p+2-k}$, where we have used $\gamma_x^2=\mathbf{1}$. This
is either $\sim \gamma_y\gamma_z$ for odd $k$ or $\sim \mathbf{1}$
for even $k$. The Clifford algebra of the Majorana operators,
$\{\gamma_\alpha,\gamma_\beta\}=2\delta_{\alpha\beta}$ and
$\gamma_\alpha^\dagger=\gamma_\alpha$ for $\alpha,\beta =x,y,z$,
leads to a closed $\mathfrak{so}(3)$ algebra of the bilinears
$\gamma_\alpha\gamma_\beta$, with $\alpha \neq \beta$. This, along
with the nested commutator in Eq.~\ref{eq:heff_comm}, implies
that $\mathbf{1}$ cannot appear in this expansion, i.e., only odd $k$ terms appear 
and all the terms in the effective Hamiltonian at all orders of high
frequency expansion are proportional to $\gamma_y\gamma_z$.

In the high frequency expansion to ${\cal O}(h^2/\omega)$ (see Appendix~\ref{sec:HFE-4th-order}
for ${\cal O}(h^4/\omega^3)$ terms), we have 
\bqa
\nonumber \can^{(1)}(t)&=&-\frac{\bar{h}}{\bar{\omega}}\cos(\bar{\omega}t
+\bar{\phi})\ i\gamma_{x}\gamma_{y}-\frac{h}{\omega}\cos(\omega t+\phi)\ i\gamma_{z}\gamma_{x},\\
\nonumber\can^{(2)}(t)&=&-\frac{h\bar{h}}{2\omega\bar{\omega}}\left(\frac{\Delta
   \omega}{\omega+\bar{\omega}}\right)\cos[\left(\omega+\bar{\omega}\right)t
+\phi+\bar{\phi}]\ i\gamma_{y}\gamma_{z},\\
\mathcal{H}^{(1)}(t)&=&\frac{h\bar{h}}{2\omega\bar{\omega}}\left(\omega+\bar{\omega}\right)\sin
(\Delta \omega t +\Delta \phi)\ i\gamma_{y}\gamma_{z}.\label{eq:eff-Ham: O(1/w)-H}
\eqa
where $\Delta \phi = \phi-\bar{\phi}$. At higher orders, the only terms in $\mathcal{H}$ are 
$\sim \sin[k(\Delta\omega t+\Delta\phi)]$ with odd $k$ because, as explained before, 
only odd $k$ terms appear in the series expansion of $\mathcal{H}$. There are no terms 
$\sim \cos[k(\Delta\omega t+\Delta\phi)]$ because in the limit $\omega\rightarrow\bar{\omega}$
and $\phi\rightarrow\bar{\phi}$, the nested commutator in Eq.~\ref{eq:heff_comm} forces 
$\mathcal{H}$ to $0$. The effective Hamiltonian is thus modulated 
with the beating frequency and its odd harmonics,
and a phase which can be controlled by the phase difference between
the two drives. This tunable effective Hamiltonian will be the
basic building block of our protocol to braid the MZMs. 

\begin{figure*}[t]
\centering 
\includegraphics[width =0.95\textwidth]{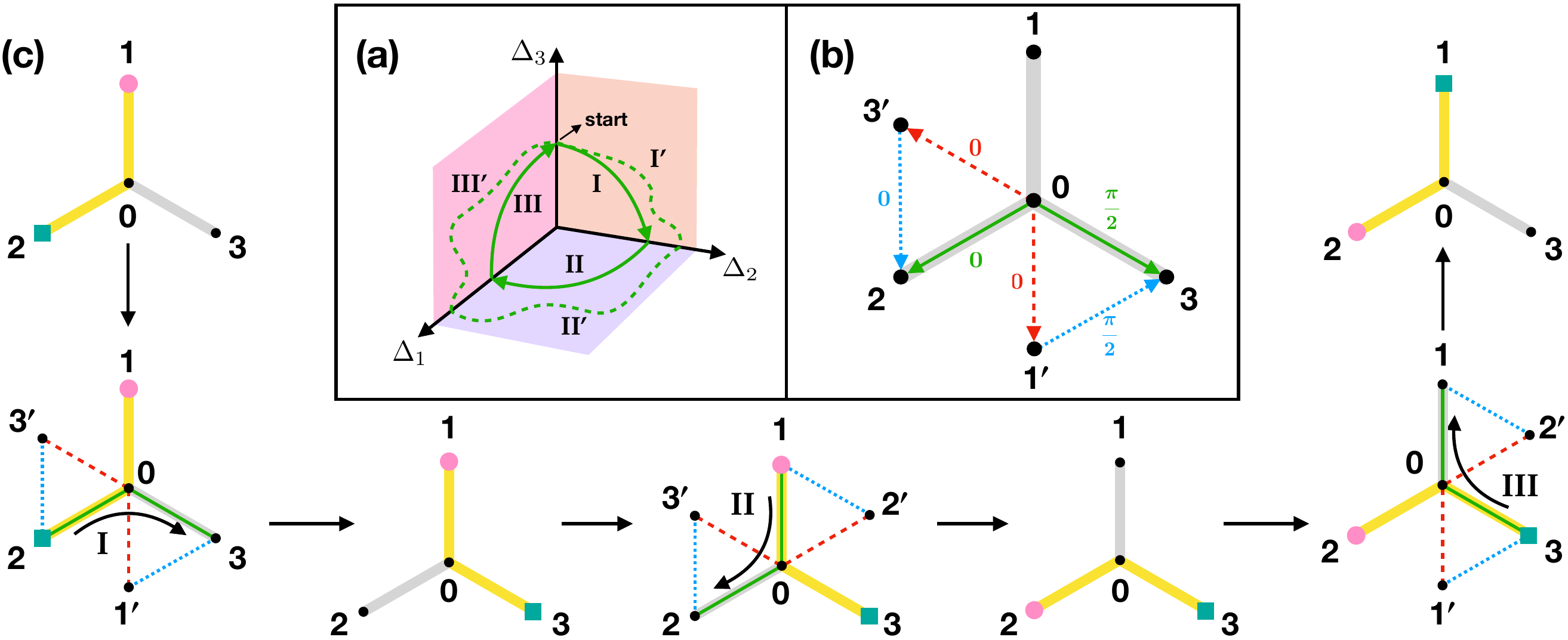}
\caption{(a) The solid green path is one possible path for braiding in the parameter space of $\Delta_1$, $\Delta_2$ and $\Delta_3$ which subtends a solid angle of $\pi/2$ at the origin. A smooth deformation of the each path segment within the corresponding plane (shown as dashed green line) still subtends the same solid angle. (b) The basic setup containing three wires (gray segments) meeting at a junction with Majorana operators $1, 2, 3$ at the ends and $0$ at the junction. Two additional operators $1'$ and $3'$ are placed for reasons explained in the text. Like before, the dashed red and dotted blue lines correspond to drives with frequencies $\bar{\omega}$ and $\omega$ respectively, and with respective phases shown. The solid green lines represent the effective low frequency couplings. The relative phases tell us this represents step $\mathbf{I}$ of the path of our protocol. (c) The three steps of braiding. The yellow segments (with unpaired MZMs at the ends) are in topological phase and the gray segments are in trivial phase. Initially, the unpaired MZMs [cyan (square) and pink (circle)], are at 1 and 2 respectively. After performing the three steps, $\mathbf{I}$, $\mathbf{II}$ and $\mathbf{III}$, of braiding, the cyan (square) and pink (circle) MZMs end up at 2 and 1 respectively, resulting in an anticlockwise braiding.}
\label{fig:braid}
\end{figure*}

\section{Protocol for braidng}

We now describe the detailed protocol to braid Majorana fermions using
high frequency tunneling. The basic setup consists of four Majorana
fermions, $\gamma_{0,1,2,3}$, at the end of three wires (gray segments), as shown 
in Fig.~\ref{fig:braid}(b). The different wires can be
driven into topological or non-topological phases by changing the
coupling $\Delta_i$ between $\gamma_0$ and $\gamma_i$ ($i=1,2,3$). 
Such coupling can be obtained by modulating the 
chemical potential in the corresponding wire~\cite{Alicea2011}. 
In fact, an MZM can be transferred from
one end to another by suitable tuning of the couplings~\cite{Alicea2011}. The braiding
operation can then be represented as a closed path traversed by the system in
the parameter space of $\Delta_1$, $\Delta_2$ and $\Delta_3$~\cite{Karzig2016}. Two possible 
paths are shown in Fig.~\ref{fig:braid}(a). We shall consider the solid green path now. 
Starting from the point shown in figure, 
the solid green path consists of three segments in the orthogonal 
planes $\Delta_2-\Delta_3$, $\Delta_1-\Delta_2$ and $\Delta_3-\Delta_1$, 
traversed in the sequence shown in the figure,
leading to an anticlockwise braiding of MZMs at $\gamma_1$ 
and $\gamma_2$. The braiding operator for this path is 
$\exp(-\pi\gamma_1\gamma_2/4)$, where the Berry phase, $\pi/4$, 
is half the solid angle, $\pi/2$, subtended by the path at origin.

Let us consider the segment labeled $\mathbf{I}$ in 
Fig.~\ref{fig:braid}(a). Here $\Delta_2$ is tuned from zero 
to a finite value while $\Delta_3$ is tuned 
from a finite value to zero. An adiabatic passage along this path 
would shift the MZM at $\gamma_2$ to 
$\gamma_3$, as shown in Fig.~\ref{fig:braid}(c). In order to achieve 
this, we put a high frequency drive with frequency 
$\bar{\omega}$ between the Majorana operators 
$\gamma_0$ and $\gamma_{3'}$ (we shall denote such pair as ($\gamma_0$,$\gamma_{3'}$) 
from now) where $\gamma_{3'}$ is an auxiliary Majorana operator which is not part of the qubit. 
Simultaneously, we put another drive with frequency 
$\omega$ between ($\gamma_{3'}$,$\gamma_2$), which is phase-locked to the previous drive.
This is depicted in Fig.~\ref{fig:braid}(b), where the dashed red 
line and dotted blue line represent drives with frequencies $\bar{\omega}$ 
and $\omega$ respectively. (Such drives between Majorana operators at end points of different wires can 
be obtained by coupling these end points to a common quantum dot 
\cite{Karzig2015,Flensberg2011}.) We will denote this common phase to be $0$ 
and measure all other phases with respect to this common phase. The relative 
phases of both the drives are also shown in the same figure. It is easy to see
from Eq.~\ref{eq:eff-Ham: O(1/w)-H}
that the drives between ($\gamma_0$,$\gamma_{3'}$) and ($\gamma_{3'}$,$\gamma_2$) 
together yield an effective coupling between 
($\gamma_{0}$,$\gamma_2$), which is modulated as $\sim \sin \Delta\omega t$, 
shown as a solid green line in Fig.~\ref{fig:braid}(b).
Similarly, adding drives between ($\gamma_0$,$\gamma_{1'}$) and ($\gamma_{1'}$,$\gamma_3$), 
where $\gamma_{1'}$ is another auxiliary Majorana operator, 
leads to an effective coupling between ($\gamma_{0}$,$\gamma_3$) modulated 
as $\sim  \sin (\Delta\omega t +\pi/2) = \cos \Delta \omega t$. If the drives are applied for a time 
$t=\pi/2\Delta \omega$, one can achieve the goal of traversing the
path segment $\mathbf{I}$, i.e., $\Delta_2$ is driven from zero 
to a finite value, $h\bar{h}(\omega+\bar{\omega})/(2\omega\bar{\omega})$,
while $\Delta_3$ is simultaneously driven from the same finite value to zero. 
(In reality, stopping the drives exactly at $t=\pi/2\Delta \omega$ is not possible. 
If the next set of drives for $\mathbf{II}$ are not started immediately after stopping the 
set of drives for $\mathbf{I}$, all the couplings go to zero simultaneously and the 
adiabaticity is lost. This can be remedied by introducing a constant coupling between 
($\gamma_{0}$,$\gamma_2$) just \emph{before} the first set of drives are stopped and 
removing it just \emph{after} the next set of drives are started. 
This would ensure that $\Delta_2$ remains finite during the switching of drives. 
The same can be done for the other switches.)

Note that the drives between ($\gamma_0$,$\gamma_{3'}$) and ($\gamma_0$,$\gamma_{1'}$) 
might appear to generate a coupling between ($\gamma_{1'}$,$\gamma_{3'}$) at the same order 
as $\Delta_2$($\Delta_3$). However, since the frequencies and phases of these drives are same, 
there is no such coupling, irrespective of the amplitudes of these drives. 
(Even if there is any coupling between ($\gamma_{1'}$,$\gamma_{3'}$), it does not affect the 
protocol because they are not the Majorana operators we are really interested in.) 
There could be a residual coupling between ($\gamma_2$,$\gamma_3$) 
coming from higher order terms of Eq.~\ref{eq:heff_comm} which is $\sim h^4/\omega^3$, 
much smaller than the scale of $\Delta_2$ or $\Delta_3$.
Even for conservative estimates such as  $\Delta \sim 20 \text{ GHz}$, 
$\omega \sim 4 \text{ GHz}$, $h \sim 1.5 \text{ GHz}$, 
the finite value of $\Delta_2(\Delta_3)$ is $\sim h^2/\omega \sim 560 \text{ MHz}$, while the 
small residual splitting is $\sim h^4/\omega^3 \sim 80 \text{ MHz}$ which is negligible compared even 
to the practically achievable temperature $T \sim 10\text{ mK} \sim 200 \text{ MHz}$.

The detailed steps of the full braiding scheme are shown in
Fig.~\ref{fig:braid}(c). The leftmost part of the figure corresponds to
traversing the path segment $\mathbf{I}$ in the parameter space, at the end of
which the MZM $\gamma_2$ has been transferred to the
position of the operator $\gamma_3$. The lower part corresponds to
traversing the path segment $\mathbf{II}$. In this case, two drives with 
frequency $\bar{\omega}$ couple ($\gamma_0$,$\gamma_{3'}$) and 
($\gamma_0$,$\gamma_{2'}$), while two drives with frequency $\omega$ couple 
$(\gamma_{2'},\gamma_1)$ and $(\gamma_2,\gamma_{3'})$. Of these, only the last drive 
is out-of-phase. Here, $\gamma_{2'}$ is a third and last auxiliary Majorana operator. 
At the end of the sequence, the MZM at
$\gamma_1$ is moved to the position of $\gamma_2$. The rightmost part 
depicts the path segment $\mathbf{III}$, which moves the MZM at
$\gamma_3$ to $\gamma_1$ and uses an auxiliary Majorana operator
$\gamma_{2'}$. The full sequence then leads to the braiding of the
MZMs at $\gamma_1$ and $\gamma_2$ in the anticlockwise direction. The
reverse sequence of paths would lead to a clockwise braiding.

\section{Robustness of the protocol}

We shall now discuss the robustness of our protocol. Note that, in Fig.~\ref{fig:braid}(a), 
deforming any path segment within the plane in which it lies 
(e.g. deforming $\mathbf{I}$ to $\mathbf{I'}$ in $\Delta_2-\Delta_3$ plane) does not 
change the solid angle subtended at origin as long as the deformed segments lie in 
mutually orthogonal planes. This can be ensured if all segments 
meet exactly along the axes of the parameter space 
(e.g. $\mathbf{I}$ and $\mathbf{II}$ as well as 
$\mathbf{I'}$ and $\mathbf{II'}$ meet along $\Delta_2$ axis) \footnote{However, 
there is no protection against arbitrary deformation out of the 
plane even if the segments meet along the axes}. 
In our protocol, the leading order terms in $\mathcal{H}$ clearly satisfy this criterion. 
Moreover, the deformations due to higher order terms can be shown to 
satisfy this criterion using a simple algebraic argument (see Appendix~\ref{sec:braid_err_high_order}). 
So there is no braiding error from the truncation of the series. 
In a realistic drive, there is always noise. The amplitude noise has no effect because it 
deforms the path only within the planes and does not disturb the meeting point of 
different segments. Phase noise, however, can affect the meeting point. 
By construction, the effective Hamiltonian depends 
only on the relative phases of the drives. For the values of $\omega,\bar{\omega}$ 
quoted before, it is practically easy to keep the relative phase fixed on time scales 
$\sim1/\Delta\omega$ if the drives are obtained by modulating a single source. 
So our protocol is immune to various sources of noise in the drives. 
\begin{figure}[t]
\centering 
\includegraphics[width =0.45\textwidth]{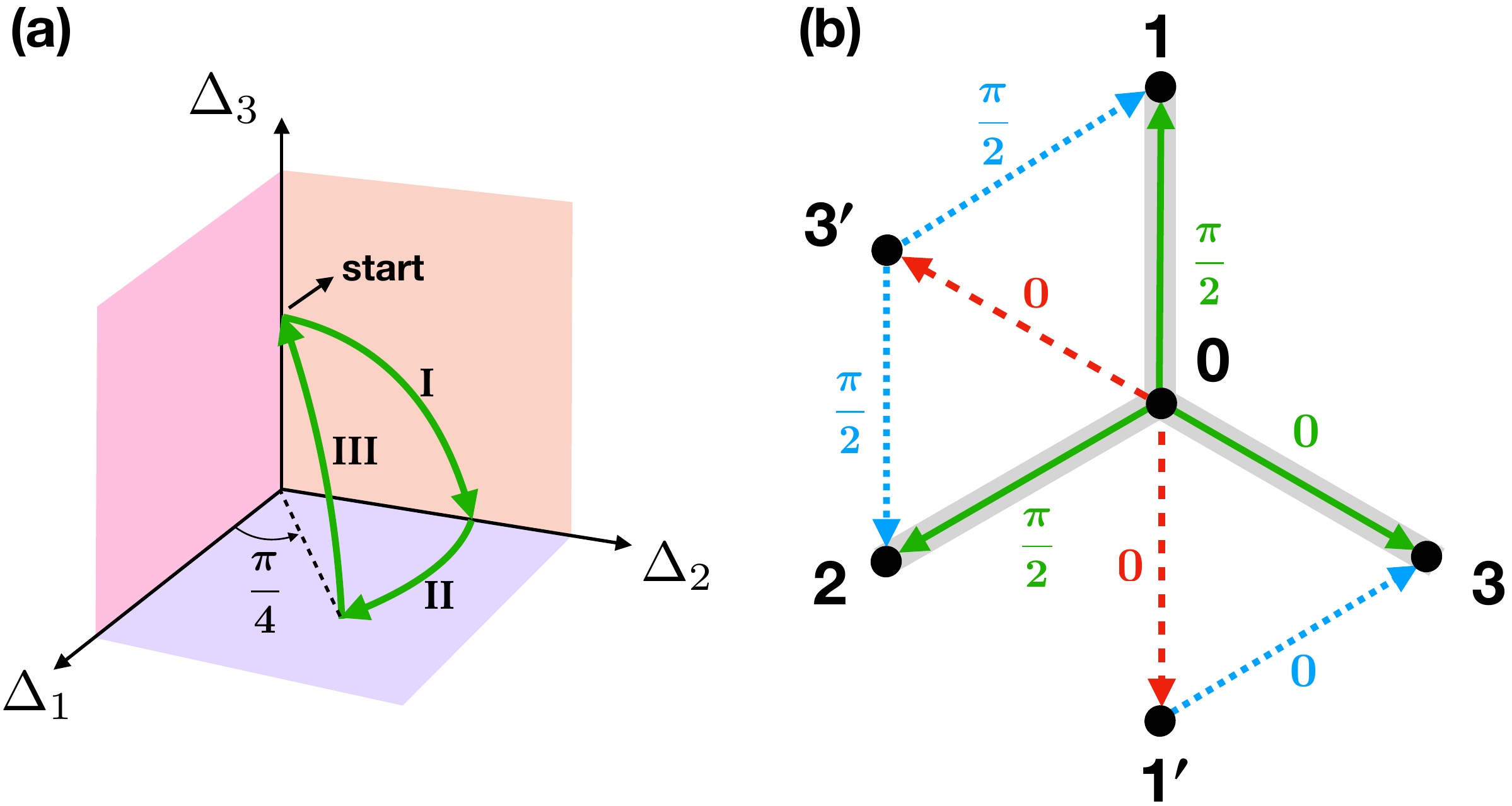}
\caption{(a) The solid green path subtends a solid angle of $\pi/4$ and hence corresponds to a phase gate. The segments $\mathbf{I}$ and $\mathbf{II}$ are both generated as in the braiding protocol explained above. The segment $\mathbf{III}$ is a topologically non-protected path because all the parameters are non-zero at the same time. (b) The setup of drives which generates the path segment $\mathbf{III}$ with usual meanings for the colors of the arrows and with respective phases shown.}
\label{fig:phasegate}
\end{figure}

\section{Protocol for phase gate}

We can extend our protocol to simulate a \emph{phase gate} through high frequency virtual tunneling by choosing suitable drives. The only difference between a phase gate and a braiding operator is that the solid angle subtended by the former is half the solid angle subtended by the latter. So the Berry phase of a phase gate is $\pi/8$ which is why it is also called a $\pi/8$-gate. Unlike the braiding operator, the phase gate is not topologically protected. One possible path for the phase gate is shown in Fig.~\ref{fig:phasegate}(a). Segment $\mathbf{I}$ is the same as in braiding and hence can be generated using the above braiding protocol. Similarly, segment $\mathbf{II}$ is half of the corresponding segment in braiding and the same protocol works by stopping the drives at $t=\pi/4\Delta\omega$ instead of $\pi/2\Delta\omega$. However, segment $\mathbf{III}$ is the infamous topologically non-protected path because all the three parameters are non-zero at the same time. Moreover, to get the correct solid angle, in addition to $\mathbf{I}$ and $\mathbf{II}$ being in the respective planes, $\mathbf{III}$ must be in the plane passing through $\Delta_3$ axis and making an angle of $\pi/4$ with the remaining axes as shown in Fig.~\ref{fig:phasegate}(a). The setup of drives shown in Fig.~\ref{fig:phasegate}(b) can generate $\mathbf{III}$ with all the desired properties. As before, the drives between ($\gamma_0$,$\gamma_{3'}$) and ($\gamma_i$,$\gamma_{3'}$) generate an effective coupling between ($\gamma_0$,$\gamma_{i}$) for $i=1,2$. Both these effective couplings are $\sim \cos\Delta\omega t$. On the other hand, the coupling between ($\gamma_0$,$\gamma_{3}$), generated by the drives between ($\gamma_{1'}$,$\gamma_{0}$) and ($\gamma_{1'}$,$\gamma_{3}$), is $\sim \sin\Delta\omega t$. This time dependence corresponds to segment $\mathbf{III}$. Note that there are no static couplings in this case between ($\gamma_1$,$\gamma_{2}$) and ($\gamma_{1'}$,$\gamma_{3'}$) irrespective of the amplitudes of the drives which generate them because both sets of drives are phase-locked.

As explained before, our protocol is immune to phase noise in the drives. Although amplitude noise did not matter in the case of braiding (because of topological protection within the planes), it does matter in the case of phase gate. It is easy to see that the Berry phase is dependent only on the ratio $h_1/h_2$, where $h_i$ is the amplitude of the drive between ($\gamma_i$,$\gamma_{3'}$), and is exactly $\pi/8$ only when $h_1/h_2=1$. So any noise in the amplitudes will not change the Berry phase as long as the amplitude noises are correlated, which is expected for phase-locked drives derived from the same source. However, if there is a systematic error, say $h_1/h_2=1+\delta$ for small $\delta$, the error in Berry phase is proportional to $\delta$ which could be significant. It was shown in~\cite{Karzig2016} that this systematic error can be reduced considerably by appropriate choice of the path in the parameter space. Our protocol can be adapted to this path and hence we can reduce any systematic errors in the Berry phase. It could be argued that the above reasoning holds even for a simple low frequency driving of the couplings along the path. But such a protocol can be affected by (absolute) phase noise in the drives unlike our protocol which depends only on the relative phases.

\section{Conclusion}

In this paper, we proposed a  way of braiding MZMs using high frequency virtual tunneling between the ends of Majorana nanowires in a tri-junction. The high frequency drives lead to an effective coarse-grained low frequency dynamics, which implements the braiding operation. The protocol is immune to amplitude noise in the drives and depends only on relative phase between drives, which can be controlled using standard phase locking techniques. We extend our protocol to show how phase gates can be implemented within a similar setup.

\begin{acknowledgments}
The authors acknowledge the use of computational facilities at
Department of Theoretical Physics, TIFR Mumbai. The authors acknowledge 
helpful discussions with Rajamani Vijayaraghavan, Bernard van Heck and Jay Deep Sau.
\end{acknowledgments}

\appendix
\section{\label{sec:HFE-4th-order}High Frequency Expansion up to $\mathcal{O}(h^4/\omega^3)$}
In the main text, we gave the expression for effective Hamiltonian to leading order $\mathcal{O}(h^2/\omega)$ in Eq.~\ref{eq:eff-Ham: O(1/w)-H}. Here, we give the derivation of these expressions up to $\mathcal{O}(h^4/\omega^3)$. The original Hamiltonian given in Eq.~\ref{eq:majorana_drive_ham} in the main text is 
\bqa
H\left(t\right)=\bar{h}\sin\left(\bar{\omega}t+\bar{\phi}\right)\ i\gamma_{x}\gamma_{y}+h\sin\left(\omega t+\phi\right)\ i\gamma_{z}\gamma_{x}.\nonumber\\
\label{eq:eff-Ham: original-Ham}
\eqa
Using the nested commutator of Eq.~\ref{eq:heff_comm} in the main text, up to $\mathcal{O}(h)$, we have 
\bqa
\mathcal{H}^{(0)}=H-\dot{\mathcal{S}}^{\left(1\right)}&=&\bar{h}\sin\left(\bar{\omega}t+\bar{\phi}\right)\ i\gamma_{x}\gamma_{y}\\
	&+&h\sin\left(\omega t+\phi\right)\ i\gamma_{z}\gamma_{x}-\dot{\mathcal{S}}^{\left(1\right)}.\nonumber
\label{eq:eff-Ham: O(1)-eqn}
\eqa
Since both the terms are of high frequency, we choose $\mathcal{S}^{(1)}$ to be 
\bqa
\mathcal{S}^{(1)}(t)=-\frac{\bar{h}}{\bar{\omega}}\cos\left(\bar{\omega}t+\bar{\phi}\right)\ i\gamma_{x}\gamma_{y}-\frac{h}{\omega}\cos\left(\omega t+\phi\right)\ i\gamma_{z}\gamma_{x}.\nonumber\\
\label{eq:eff-Ham: O(1)-S1}
\eqa
Therefore, to $\mathcal{O}(h)$, the effective Hamiltonian is 
\bqa
\mathcal{H}^{\left(0\right)}\left(t\right)=0.
\label{eq:eff-Ham: O(1)-Ham}
\eqa
Similarly, at $\mathcal{O}(h^{2}/\omega)$, we have 
\bqa
\mathcal{H}^{(1)} & =&\left[i\mathcal{S}^{\left(1\right)},\left(H-\frac{\dot{\mathcal{S}}^{\left(1\right)}}{2}\right)\right]-\dot{\mathcal{S}}^{(2)}\nonumber\\ & =&\frac{h\bar{h}}{2\omega\bar{\omega}}\left(\omega+\bar{\omega}\right)\sin\left[\left(\omega-\bar{\omega}\right)t+\phi-\bar{\phi}\right]\ i\gamma_{y}\gamma_{z} \\
	&+&\frac{h\bar{h}}{2\omega\bar{\omega}}\left(\omega-\bar{\omega}\right)\sin\left[\left(\omega+\bar{\omega}\right)t+\phi+\bar{\phi}\right]\ i\gamma_{y}\gamma_{z}-\dot{\mathcal{S}}^{(2)}.\nonumber
\label{eq:eff-Ham: O(1/w)-eqn}
\eqa
To cancel only the high frequency terms, we choose $\mathcal{S}^{(2)}$ to be 
\bqa
\mathcal{S}^{(2)}(t)=-\frac{h\bar{h}}{2\omega\bar{\omega}}\left(\frac{\omega-\bar{\omega}}{\omega+\bar{\omega}}\right)\cos\left[\left(\omega+\bar{\omega}\right)t+\phi+\bar{\phi}\right]\ i\gamma_{y}\gamma_{z}.\nonumber\\
\label{eq:eff-Ham: O(1/w)-S2}
\eqa
Therefore, at $\mathcal{O}(h^{2}/\omega)$, the effective Hamiltonian is 
\bqa
\mathcal{H}^{(1)}(t)=\frac{h\bar{h}}{2\omega\bar{\omega}}\left(\omega+\bar{\omega}\right)\sin\left[\left(\omega-\bar{\omega}\right)t+\phi-\bar{\phi}\right]\ i\gamma_{y}\gamma_{z}.\nonumber\\
\label{eq:eff-Ham: O(1/w)-Ham}
\eqa
Combining the above results, the effective Hamiltonian up to $\mathcal{O}(h^{2}/\omega)$ is 
\bqa
\mathcal{H}(t)=\frac{h\bar{h}}{2\omega\bar{\omega}}\left(\omega+\bar{\omega}\right)\sin\left(\Delta\omega t+\Delta\phi\right)\ i\gamma_{y}\gamma_{z}+O\left(\frac{h^{4}}{\omega^{3}}\right).\nonumber\\
\label{eq:eff-Ham: tot-eff-Ham}
\eqa
%
%
%
As we have shown in main text, $\mathcal{H}^{(n)}$ is non-zero only for odd $n$. Hence, the next non-zero term in $\mathcal{H}$ comes at third order. Continuing the same procedure, we obtain the following expression for $\mathcal{H}^{(3)}$, $\can^{(3)}$ and $\can^{(4)}$,
\begin{widetext}
\bqa
\mathcal{H}^{(3)}(t) &=&\frac{h\bar{h}}{12\omega\bar{\omega}}\Bigg[\frac{h^{2}\left(10\omega^{3}+2\omega^{2}\bar{\omega}-\omega\bar{\omega}^{2}+\bar{\omega}^{3}\right)}{\omega^{2}(\omega+\bar{\omega})\left(2\omega-\bar{\omega}\right)}+\frac{\bar{h}^{2}\left(10\bar{\omega}^{3}+2\bar{\omega}^{2}\omega-\bar{\omega}\omega^{2}+\omega^{3}\right)}{\bar{\omega}^{2}(\omega+\bar{\omega})\left(2\bar{\omega}-\omega\right)}\Bigg]\sin\left(\Delta\omega t+\Delta\phi\right)\ i\gamma_{y}\gamma_{z},
\label{eq:eff-Ham: O(1/w3)-Ham}
\eqa
\bqa
\mathcal{S}^{(3)}(t) &=&\frac{h\bar{h}^{2}}{12\omega\bar{\omega}^{2}}\Bigg[\frac{(\omega-\bar{\omega})(\omega-2\bar{\omega})}{(\omega+\bar{\omega})(\omega+2\bar{\omega})}\cos[(\omega+2\bar{\omega})t+\phi+2\bar{\phi}]+\frac{(5\omega+11\bar{\omega})}{(\omega+\bar{\omega})}\cos(\omega t+\phi)\nonumber\\
	& &+\frac{4(\omega+\bar{\omega})}{(\omega-2\bar{\omega})}\cos[\left(2\bar{\omega}-\omega\right)t+2\bar{\phi}-\phi]\Bigg]\ i\gamma_{z}\gamma_{x}\nonumber\\
	& &+\frac{h^{2}\bar{h}}{12\omega^{2}\bar{\omega}}\Bigg[\frac{(\bar{\omega}-\omega)(\bar{\omega}-2\omega)}{(\omega+\bar{\omega})(\bar{\omega}+2\omega)}\cos[(\bar{\omega}+2\omega)t+\bar{\phi}+2\phi]+\frac{(5\bar{\omega}+11\omega)}{(\omega+\bar{\omega})}\cos(\bar{\omega}t+\bar{\phi})\nonumber\\
	& &+\frac{4(\omega+\bar{\omega})}{(\bar{\omega}-2\omega)}\cos[(2\omega-\bar{\omega})t+2\phi-\bar{\phi}]\Bigg]\ i\gamma_{x}\gamma_{y},
\label{eq:eff-Ham: O(1/w2)-S3}
\eqa
\bqa
\mathcal{S}^{(4)}(t) &=&\frac{h\bar{h}}{24(\omega+\bar{\omega})}\Bigg[\left(\frac{h^{2}(\bar{\omega}^{2}+7\omega\bar{\omega}+22\omega^{2})}{\omega^{3}\bar{\omega}(\bar{\omega}+2\omega)}-\frac{\bar{h}^{2}(\omega^{2}+7\omega\bar{\omega}+22\bar{\omega}^{2})}{\omega\bar{\omega}^{3}(\omega+2\bar{\omega})}\right)\cos[(\omega+\bar{\omega})t+\phi+\bar{\phi}]\nonumber\\
	& &+\left(\frac{\bar{h}^{2}(6\bar{\omega}-\omega)(\omega+\bar{\omega})^{2}\cos[(\omega-3\bar{\omega})t+\phi-3\bar{\phi}]}{\omega\bar{\omega}^{3}(3\bar{\omega}-\omega)(2\bar{\omega}-\omega)}-\frac{h^{2}(6\omega-\bar{\omega})(\omega+\bar{\omega})^{2}\cos[(\bar{\omega}-3\omega)t+\bar{\phi}-3\phi]}{\omega^{3}\bar{\omega}(3\omega-\bar{\omega})(2\omega-\bar{\omega})}\right)\nonumber\\
	& &+\left(\frac{2\bar{h}^{2}(\bar{\omega}-\omega)\cos[(\omega+3\bar{\omega})t+\phi+3\bar{\phi}]}{\bar{\omega}^{2}(\omega+2\bar{\omega})(\omega+3\bar{\omega})}-\frac{2h^{2}(\omega-\bar{\omega})\cos[(\bar{\omega}+3\omega)t+\bar{\phi}+3\phi]}{\omega^{2}(\bar{\omega}+2\omega)(3\omega+\bar{\omega})}\right)\Bigg]\ i\gamma_{y}\gamma_{z}.
\label{eq:eff-Ham: O(1/w3)-S4}
\eqa
\end{widetext}

\section{\label{sec:braid_err_high_order}No contributions to braiding error at higher orders}
At the leading order, we can see that the criteria for protection described in the main text are satisfied in our protocol. Now, we shall argue that even at higher orders, these criteria are still satisfied. For this, we just have to know the form of $\Delta_i$ in $\mathcal{H}$ at all orders. A similar question was answered in the main text in the simpler case of three Majorana operators in the building block. We repeat the argument here for completeness. We know that $\mathcal{H}^{(n)}$ is non-zero only for odd $n$. At order $2p+1$ (p is non-negative integer), $\mathcal{H}^{(2p+1)}$ is a product of terms $\sim (\gamma_x\gamma_y)^k (\gamma_x\gamma_z)^{2p+2-k}\sim \gamma_y^k\gamma_z^{2p+2-k}$, where we have used $\gamma_x^2=1$. This is either $\sim \gamma_y\gamma_z$ for odd $k$ or $\sim 1$ for even $k$. The Clifford algebra of the Majorana operators, $\{\gamma_\alpha,\gamma_\beta\}=2\delta_{\alpha\beta}$ and $\gamma_\alpha^\dagger=\gamma_\alpha$ for $\alpha,\beta =x,y,z$, leads to a closed $\mathfrak{so}(3)$ algebra of the bilinears $\gamma_\alpha\gamma_\beta$, with $\alpha \neq \beta$. This, together with the nested commutator form in Eq.~\ref{eq:heff_comm} in the main text, implies that identity $1$ does not appear in this expansion, i.e., only odd $k$ terms appear and hence all the terms in the effective Hamiltonian obtained at different orders of high frequency expansion are proportional to $\gamma_y\gamma_z$. Moreover, at higher orders, the only terms in $\mathcal{H}$ are $\sim \sin[k(\Delta\omega t+\Delta\phi)]$ with odd $k$ because, as explained above, only odd $k$ terms appear in the series expansion of $\mathcal{H}$. There are no terms $\sim \cos[k(\Delta\omega t+\Delta\phi)]$ because in the limit $\omega\rightarrow\bar{\omega}$ and $\phi\rightarrow\bar{\phi}$, the nested commutator in Eq.~\ref{eq:heff_comm} in the main text forces $\mathcal{H}$ to $0$. These facts can be verified up to $\mathcal{O}(h^4/\omega^3)$ by the expressions for $\mathcal{H}^{(1)}$ and $\mathcal{H}^{(3)}$ in Eq.~\ref{eq:eff-Ham: O(1/w)-Ham} and Eq.~\ref{eq:eff-Ham: O(1/w3)-Ham} respectively.

Let us now give a similar argument for segment $\mathbf{I}$ of Fig.~\ref{fig:braid}(a). We expect the effective Hamiltonian to be of the form $\Delta_2\ i\gamma_0\gamma_2+\Delta_3\ i\gamma_0\gamma_3$, where $\Delta_2$ starts from $0$ and reaches a finite value while $\Delta_3$ starts from a finite value and goes to $0$ to satisfy the criteria. It follows from the nested commutator in Eq.~2 in the main text that, at any order in $\mathcal{H}$, the generic coupling of $\gamma_0\gamma_2$ comes from 
\bqa
&\sim& (\gamma_0\gamma_{3'})^{k_1}(\gamma_{3'}\gamma_2)^{k_2}(\gamma_0\gamma_{1'})^{k_3}(\gamma_{1'}\gamma_3)^{k_4}\nonumber\\
	&=& \gamma_0^{k_1+k_3}\gamma_2^{k_2}\gamma_3^{k_4}\gamma_{1'}^{k_3+k_4}\gamma_{3'}^{k_1+k_2},\nonumber
\eqa
where $k_1,k_2,k_3$ and $k_4$ are non-negative integers such that $k_1+k_2+k_3+k_4$ is even because $\mathcal{H}^{(n)}$ is non-zero only for odd $n$.  The above form comes from the four drive terms shown as dashed red and dotted blue arrows in Fig.~\ref{fig:braid}(b). Clearly, to get a $\gamma_0\gamma_2$ term, we need $k_1+k_3$ and $k_2$ to be odd, and $k_1+k_2$, $k_3+k_4$ and $k_4$ to be even. Since $k_4$ is \emph{even}, the drive term with a relative phase of $\pi/2$ (the $\gamma_{1'}\gamma_3$ term) appears \emph{even} number of times. Hence, the phase of the coupling of $\gamma_0\gamma_2$ is always an even multiple of $\pi/2$. On the other hand, a similar analysis for $\gamma_0\gamma_3$ shows that $k_1+k_3$ and $k_4$ must be odd, and $k_1+k_2$, $k_3+k_4$ and $k_2$ must be even. Since $k_4$ is \emph{odd} in this case, the drive term with a relative phase of $\pi/2$ appears \emph{odd} number of times. Hence, the phase of the coupling of $\gamma_0\gamma_3$ is always an odd multiple of $\pi/2$. Therefore, $\Delta_2 \sim \sin(k\Delta\omega t + \text{``even''}\pi/2) \sim \sin(k\Delta\omega t)$, where $k$ is odd because $k_1+k_3$ above is odd, and similarly, $\Delta_3 \sim \sin(k\Delta\omega t + \text{``odd''}\pi/2) \sim \cos(k\Delta\omega t)$, where $k$ is odd. These forms ensure that $\Delta_3$ starts from a finite value at $t=0$ and goes to $0$ at $t=\pi/2\Delta\omega$ while $\Delta_2$ starts from $0$ at $t=0$ and reaches a finite value at $t=\pi/2\Delta\omega$, thereby satisfying the criteria mentioned in the main text for protection.
  
\bibliographystyle{apsrev4-1}
\bibliography{Majorana_Dyn}

\end{document}